\documentclass[preprint,12pt,authoryear]{elsarticle}
\usepackage{amssymb}
\usepackage{graphics,epsfig,  amsmath}
\usepackage{latexsym,color, amsthm,fullpage}

\journal{Annals of Applied Probability}

\newtheorem{theorem}{Theorem}
\newtheorem{lemma}{Lemma}

\def\god{\stackrel{d}{\longrightarrow}}

\def\bi{\begin{itemize}}
\def\ei{\end{itemize}}

\def\ben{\begin{enumerate}}
\def\een{\end{enumerate}}

\newcommand*\xbar[1]{%
  \hbox{%
    \vbox{%
      \hrule height 0.5pt 
      \kern0.3ex
      \hbox{%
        \kern-0.1em
        \ensuremath{#1}%
        \kern-0.1em
      }%
    }%
  }%
} 

\begin{document}
\begin{frontmatter}
\title{Convergence of  empirical distributions  
in an\\  interpretation of  quantum mechanics}
\author{Ian W. McKeague and Bruce Levin}

\address{Department of Biostatistics, Columbia University}  
\begin{abstract}  
From its beginning, there have been attempts by physicists to formulate quantum mechanics without requiring the use of wave functions.   An interesting recent approach takes the point of view that quantum effects arise solely from the interaction of finitely many classical ``worlds."  The wave function is then recovered (as a secondary object) from observations of  particles in these worlds, without knowing the world from which any particular observation originates. Hall, Deckert and Wiseman [{\it Physical Review} X {\bf 4} (2014) 041013] have introduced an explicit many-interacting-worlds harmonic oscillator model to provide support for this approach. In this note we provide a proof of their claim that the particle configuration is asymptotically Gaussian, thus matching the stationary ground-state solution of Schr\"odinger's equation when the number of worlds goes to infinity.  We also construct a Markov chain based on resampling from the  particle configuration and show that it converges to an Ornstein--Uhlenbeck process, matching the time-dependent solution as well.
\end{abstract}


\end{frontmatter}

\newenvironment{packenum}{
\begin{enumerate}
  \setlength{\itemsep}{2pt}
  \setlength{\parskip}{2pt}
  \setlength{\parsep}{0pt}
}{\end{enumerate}}

\renewcommand\baselinestretch{1.5}

\numberwithin{equation}{section}

\section{Introduction}

Let  $x_1,\ldots, x_N$ be a finite sequence of real numbers satisfying the recursion relation 
\begin{equation}x_{n+1}=x_n-  {1\over x_1+\ldots +x_n}.  
\label{rec}
\end{equation}
In this note we show that for a certain class of solutions (monotonic with zero-mean), the  empirical distribution of the $x_n$ converges to  standard Gaussian as $N\to \infty$.  We also construct a simple Markov chain based on resampling from this empirical distribution and show that it converges to an Ornstein--Uhlenbeck process.
 
 \cite{Hal2014} derived the recursion relation (\ref{rec}) via Hamiltonian mechanics and used it to justify a novel interpretation of quantum mechanics.  The solution they considered represents the stationary ground-state configuration of a harmonic oscillator in $N$  ``worlds," where $x_n$ is the location (expressed in dimensionless units) of a particle in the $n$th world. The particles  behave classically (deterministically in accordance with Newtonian mechanics) within  each world, and there is a mutually repulsive force  between particles in adjacent worlds.  Observers  have access to  draws from the empirical distribution  $$\mathbb{P}_N(A)={\#\{n\colon x_n\in A\}\over N}$$  
for any Borel set $A\subset \mathbb{R}$,  but  do not know the world from which any observation originates due to their ignorance as to which world they occupy. In statistical language, Efron's nonparametric bootstrap can be used by observers (to obtain draws with replacement from the whole configuration $\{x_1,\ldots, x_N\}$),  but they are unable to identify any particular  $x_n$.  
  
 \cite{Hal2014} discovered that $\mathbb{P}_N$ is approximately Gaussian, thus corresponding  to the stationary ground-state solution of  Schr\"odinger's equation for the wave function of a particle in a parabolic potential well, and  furnishing a many-interacting-worlds interpretation of this wave function. They provided convincing numerical evidence that the Gaussian approximation is accurate when  $N=11$, a case in which the recursion relation admits an exact solution. As far as we know, however,  a formal proof of convergence is not yet available.   
\cite{Seb2014} independently proposed a similar many-interacting-worlds interpretation, called Newtonian quantum mechanics, although no explicit example was provided. Our interest in studying the explicit  model (\ref{rec}) is that rigorous investigation of its limiting behavior becomes feasible.   Both \cite{Hal2014} and  \cite{Seb2014} noted  the ontological difficulty of a continuum of worlds, a feature of an earlier but closely related hydrodynamical approach due to \cite{Hol2005},  \cite{Poi2010} and \cite{Sch2012}.    

The motivation for the recursion  (\ref{rec}) given by \cite{Hal2014} was to explore explicitly the consequences of replacing the continuum of fluid elements in the Holland--Poirier theory by a ``huge" but nevertheless finite number of interacting worlds. Yet their approach raises the question of whether such a discrete model has a stable solution when the number of worlds becomes large.  A  formal way to address this question is to establish the convergence of  $\mathbb{P}_N$ under suitable conditions.  The problem is  non-trivial, however, because explicit solutions of the recursion are only available for small values of $N$, and numerical methods are useful only for exploratory purposes.  Nevertheless, we are able to establish our result using only standard methods of distribution theory, and most crucially the Helly selection theorem.  In addition, by making use of Stein's method, we are able to find a rate of convergence.  We further construct a Markov chain based on bootstrap resampling from $\mathbb{P}_N$ and show that it converges to the Ornstein--Uhlenbeck process corresponding to the full (time-dependent) ground-state solution of Schr\"odinger's equation in this setting.  

Our main results are collected in Section 2, and  their proofs are in Section 3.  For general background on parallel-world theories in quantum physics, we refer the interested reader to the book of  \cite{Gre2011}.  For the convenience of the reader, at the end of Section 3 we have provided  \cite{Hal2014}'s derivation of (\ref{rec}) as representing a ground state solution of the many-interacting-worlds Hamiltonian.

\section{Main results}  

Our main result is that   $\mathbb{P}_N$ 
has a standard Gaussian limit  for  monotonic zero-mean  solutions $\{x_1,\ldots, x_N\}$ to the recursion  (\ref{rec}). Monotonicity  and zero-mean (along with the recursion) are necessary conditions for a ground-state solution of the many-interacting-worlds Hamiltonian, so our result establishes in full generality the  normal approximation claimed by  \cite{Hal2014}.
We also show that such solutions to the recursion exist and are unique for each $N\ge 3$ (Lemma \ref{lem1} in Section 3).  The solutions are indexed by $N$, and  for clarity in the proofs we  will  write $x_n$ as $x_{N,n}$.  For now, though, we  suppress the dependence on $N$.   Our main result is as follows.

\begin{theorem} 
\label{thm}  The unique monotonic zero-mean solution $\{ x_n, n=1,\ldots, N\}$ of the recursion relation (\ref{rec}) satisfies $\mathbb{P}_N\god {\cal N}(0,1)$ as $N\to \infty$.
\end{theorem}

\subsection*{Remarks}
\ben
\item 
Our proof  of Theorem \ref{thm} will proceed by showing that $\mathbb{P}_N$ is close in distribution to a certain piecewise-constant density $y_N(x)$, which in turn is shown to converge pointwise to the ${\cal N}(0,1)$ density.  Further, we will construct a coupling between two random variables $X_N\sim \mathbb{P}_N$ and $\tilde X_N \sim y_N$ such that $|X_N-\tilde X_N|\to 0$  almost surely, so the result  will then follow from Slutsky's lemma.    An illustration of $y_N(x)$ is given in Fig.\ 1.

\item Stein's method, as often used for studying normal approximations to sums of independent random variables \cite[see][]{che2010}, is   applicable in our setting and gives   insight into the rate of convergence. We will discuss this  approach following the proof.    Stein's method becomes particularly easy to apply in our setting because  $y_N(x)$ is the so-called zero-bias density of $\mathbb{P}_N$.  

\item Let $w_1,\ldots, w_N$ satisfy the more general recursion relation
$$w_{n+1}=w_n-  {\sigma^2\over w_1+\ldots +w_n},$$ where $\sigma^2>0$.  The scaled sequence  $x_n=w_n/\sigma $ satisfies (\ref{rec}), so Theorem \ref{thm} applies and the empirical distribution of $\{w_1,\ldots, w_N\}$  converges to ${\cal N}(0,\sigma^2)$.  It is  striking that the variance $\sigma^2$, rather than the standard deviation, appears linearly in the  recursion for $w_n$. 
 For the harmonic oscillator  studied by \cite{Hal2014},  $\sigma^2=\hbar/(2m\omega)$, where $\hbar$ is the reduced Planck constant, $m$ is the mass of the particle, and $\omega$ is the angular frequency.

\item
Monotonicity of the solution to the recursion may not  be necessary, even though our proof of Theorem \ref{thm} relies on it.  We have found from numerical experiments that {\it non}-monotonic  solutions can exist  with $\mathbb{P}_N$  indistinguishable from standard Gaussian. 
 From the physical point of view, however, monotonicity  is an essential requirement: the ordering of the particles is always preserved by the repulsive nature of the interaction between   worlds  (\citealt{Hal2014}, Section III). 


\item Theorem \ref{thm} is equivalent to the statement that if  $n=n(N)\to \infty$ with $n/N\to \alpha$ for some $0<\alpha<1$, then $x_n$ converges to the upper-$\alpha$-quantile of ${\cal N}(0,1)$. This implies a simple recursion approximation for {\it intermediate} normal quantile.

\een

\begin{figure}[!ht]
\begin{center}
      \includegraphics[scale=.4]{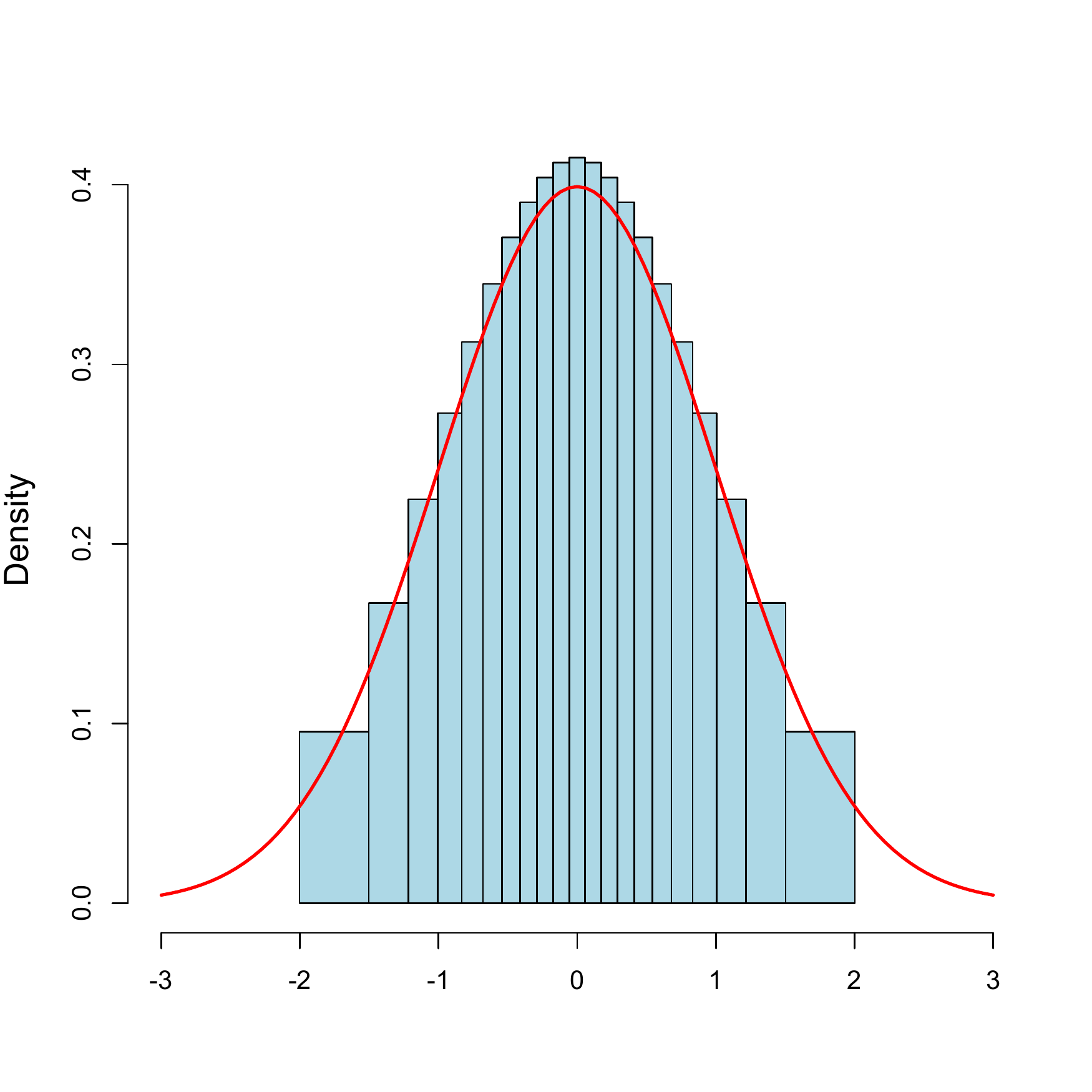}
 \caption{Harmonic oscillator ground state configuration for $N=22$ worlds compared with the ${\cal N}(0,1)$ density. The histogram is the $\mathbb{P}_N$-zero-bias density $y_N(x)$, and the breaks in the histogram are successive $x_n$.}
   \label{hist}
 \end{center}
   \end{figure}

\section*{Numerical example}

The recursion can be rapidly iterated, but to generate an exact  monotonic zero-mean solution,  $x_1$ needs to be known.  A ``randomly chosen" initial point will not lead to a solution, and the system is non-robust to the choice of $x_1$, which parallels the physics in the sense that explicit  ground-state solutions of the many-interacting-worlds Hamiltonian are not available.  

We consider an example with $N=22$ worlds, and use the following trial-and-error approach to obtain an approximate solution. Given the proximity of  $x_n$ to  normal quantiles  noted above, the $1/N$-upper-quantile of ${\cal N}(0,1)$ might be considered as a suitable initial choice of $x_1$. From our numerical experiments, however, we have found that the $1/(2N)$-upper-quantile (denoted $q_N$) is much more accurate; the normal approximation is poor in the extreme tail of $\mathbb{P}_N$ (i.e., more extreme than $\pm x_1$), and the scaling by 2 compensates well for this.  A search over a fine grid in a small neighborhood of $q_N$ then quickly yields a very accurate solution by minimizing $|x_1+x_N|$.  In this example,  $q_N = 2.0004$ and the best approximation is  $x_1=2.0025$ (to 4 decimal places).   

Fig.\ \ref{hist} displays the $\mathbb{P}_N$-zero-bias density $y_N(x)$ having mass $1/(N-1) $ uniformly distributed over the intervals between successive $x_n$, along with the  ${\cal N}(0,1)$ density.  The approximation is remarkably good except around zero and in the extreme tails.

\section*{General solutions to the quantum harmonic oscillator}
Using an approach  to quantum mechanics pioneered by Edward Nelson, it can be shown that the full ground state solution of Schr\"odinger's equation for a harmonic oscillator can be represented in terms of the distribution of an Ornstein--Uhlenbeck (OU) process.  Moreover, the complete family of solutions can  be represented by adding this ground-state process to all solutions of the classical harmonic oscillator; see, e.g., 
\cite{Pau2013}, pages 122--124.  The  limit in Theorem \ref{thm} refers to the stationary distribution of this ground-state OU process, but  it is also possible to construct a many-interacting-worlds  approximation to the  OU process itself.  This can be done in terms of simple random samples from $\mathbb{P}_N$ that evolve as a Markov chain, as we now explain.  

Let $\{Z_1,\ldots, Z_m\}$ be an independent random sample of size $m$ from ${\cal N}(0,1)$,  corresponding  to $m$ draws from $\mathbb{P}_N$ in the limit as $N\to\infty$ (by Theorem \ref{thm}).  Let  $Y_0=\sum_{i=1}^m Z_i$, and let $Y_1$ be  obtained from $Y_0$ by replacing a randomly selected $Z_i$ by an independent draw from ${\cal N}(0,1)$. By iterating this ``random single replacement"   mechanism (cf.\ sequential bootstrap), we obtain a stationary Markov chain of samples of size $m$,  and an autoregressive Gaussian time series $Y_k$ satisfying 
$$Y_k =(1-\lambda)Y_{k-1} +\epsilon_k, \ \ k=1,2, \ldots $$
where $\lambda=\lambda_m=1/m$, and the innovations $\epsilon_k\sim {\cal N}(0,2-\lambda)$ are independent of each other and of past values of the time series  \cite[cf.][pages 22--25]{che2010}.  Construct a rescaled version of the time series as a  random element of the Skorohod space $D[0,\infty)$ by setting $$X_t^{(m)}=Y_{[mt]}/\sqrt m, \ \  t\ge 0,$$ 
where $[\cdot ]$ is the integer part. Using a result of 
\cite{Phi1987} concerning  first-order autoregressions with a root near unity, we can show that $  X_t^{(m)}$ converges in distribution as $m\to\infty$ to the (stationary) OU process $X_t$ that satisfies the stochastic differential equation
$$dX_t=-X_t\, dt +\sqrt 2\, dW_t, \ \ t\ge 0,$$
where $W_t$ is a standard Wiener process and $X_0\sim {\cal N}(0,1)$.  
It suffices to consider the  time series
$$y_k=Y_k/\sqrt{1-\lambda/2}, \ \ k=1,2, \ldots,$$ which has iid-${\cal N}(0,2)$ innovations  and autoregressive parameter $a=(1-\lambda)/\sqrt{1-\lambda/2}$.  Setting $a=e^{c/T}$ where $c=-1$ and $T=T_m\to \infty$ to match Phillips's  notation, we have   $T/ m\to 1$, so his Lemma 1 (a) gives that the process $y_{[mt]}/\sqrt m$ converges in distribution  to the OU process $X_t$, as required.  

\medskip
We expect the same limit result if the Markov chain consists of random samples of size $m$ from $\mathbb{P}_N$ (that evolve by the same mechanism), provided  $N$ and $m$ simultaneously tend to infinity.  The proof of such a general result would be difficult, however, as it would involve extending the above argument to  time series in which the innovations depend on $m$ and $N$ and that are no longer  independent.  Nevertheless, we can show that such a result holds provided $m=m_N\to \infty$ slowly enough, as follows. 

\begin{theorem} 
\label{thm2} Suppose the conditions of Theorem \ref{thm} hold.   Let $Y_{N,k}$ be the  sum of values at time $k$ of the stationary Markov chain of samples of size $m$ generated by the mechanism of random single replacement  from $\mathbb{P}_N$.  Then, if   $m=m_N\to \infty$  and $m=o({\log N})^{1/3}$ the rescaled process    $$\xbar X_{t}=Y_{N,[mt]}/\sqrt m, \ \  t \ge 0$$ converges in distribution on $D[0,\infty)$ to the OU process $X_t$.
\end{theorem}

\section{Proofs } 

Before proceeding to  the proof of Theorem \ref{thm}, we state a lemma (to be proved later) that gives the key properties needed to establish the theorem, and also establishes the existence of a {\it unique} solution to the recursion relation that satisfies the conditions of the theorem.  Throughout we implicitly assume $N\ge 3$.  We will also make extensive use of the notion of {\it zero-median} in the following sense: if $N$ is odd, then  $x_{(N+1)/2}=0$; if $N$ is even, then $x_{N/2}=-x_{(N/2) +1}$.


\begin{lemma}  
\label{lem1}
Every zero-median solution $x_1,\ldots, x_N$ of  (\ref{rec}) satisfies the following properties:
\begin{itemize}
\item[]  
\begin{itemize} 
\item[{\rm(P1)}] Zero-mean: $x_1+\ldots +x_N=0$.
\item[{\rm (P2)}]  Variance-bound:  $x_1^2+\ldots +x_N^2=N-1$.
\item[{\rm(P3)}]  Symmetry:  $x_n=-x_{N+1-n}$ for $n=1,\ldots, N$.
\end{itemize}
\end{itemize}
Further, there is a unique  solution $x_1,\ldots, x_N$ of  (\ref{rec}) 
such that {\rm (P1)} and  
\begin{itemize}
\item[] 
\begin{itemize} 
\item[{\rm(P4)}]  Strictly decreasing:  $x_1>\ldots >x_N$
\end{itemize}
\end{itemize}
hold.  This   solution  has the zero-median property, and thus also satisfies {\rm (P2)} and {\rm (P3)}.
\end{lemma}
\medskip

Without loss of generality we can assume that (P4) holds, since if we start with an increasing zero-median solution, reversing the order of the solution provides a decreasing solution by (P3), and any monotonic solution is strictly monotonic. The  dependence on $N$ is now made explicit:  write $x_n=x_{N,n}$, and also denote $S_{N,n}=x_{N,1} +\ldots + x_{N,n}$ for $n=1,\ldots, N$.  

\bigskip

First we show that  $x_{N,1}\ge \sqrt{\log N}/2$.  For odd  $N$, the median $x_{N,m}=0$, where  $m=m_N=(N+1)/2$, and the telescoping sum 
 $$x_{N,1} = \sum_{n=1}^{m-1} (x_{N,n}-x_{N,n+1}) >{1\over x_{N,1}} \sum_{n=1}^{m-1} {1\over n} >{\log m\over x_{N,1}}.$$
Here we used  (\ref{rec}), (P4) and  $S_{N,n} < n x_{N,1}$ for the first inequality, and  Euler's approximation to the partial sums of the harmonic series for the second inequality.  This gives
 $x_{N,1}>\sqrt{\log m}$, and a similar argument shows that the same is true for $N$ even with $m=N/2$.  The claim then  follows using the inequality $\log(N/2) >(\log N)/4$ for $N\ge 3$.
 
Now using  the symmetry property to bound $S_{N,n}$  from below by $x_{N,1}$ for  $n=1,\ldots, N-1$,   the recursion (\ref{rec}) gives the uniform bound
$$0<x_{N,n}-x_{N,n+1} =1/S_{N,n} \le 1/x_{N,1},$$  
and an upper bound on the mesh of the sequence:
\begin{equation}\delta_N\equiv \max_{n=1,\ldots,N-1}|x_{N,n}-x_{N,n+1}| \le  2/\sqrt{\log N}. 
 \label{mesh}
\end{equation} 

\medskip

Next, for $x\in \mathbb{R}$ such that $|x|< x_{N,1}$, let $n=n(x,N)$ be the unique index satisfying $x_{N,n+1}\le  x < x_{N,n}$, and define  $y_{N}(x)=S_{N,n}/(N-1)$, so from the recursion relation (\ref{rec}) we have
  \begin{equation}y_N(x) = [(N-1) (x_{N,n} -x_{N,n+1})]^{-1}.
\label{den}
\end{equation}  
Defining  $y_{N}(x)=0$ for  $|x|\ge x_{N,1}$ makes $y_N$ into a piecewise-constant density; see Fig.\ 1 for an illustration.

\medskip

We will show that   $ y_{N}(x)$ converges uniformly in $x$, although we only need pointwise convergence. 
Let $X_N$ be a random variable distributed according to  the empirical distribution $\mathbb{P}_N$ that was defined in the Introduction. Set $Y_N(x) = X_N I(X_N > x)$.  Since $y_{N}(x) = (N/(N-1))E Y_N(x)$, it suffices to consider $E Y_N(x)$.  We use  a subsequence argument. Note that $X_N$ has second moment $1-1/N$ by (P2), so it is bounded in probability (tight).  Thus, by the Helly selection theorem, there is a subsequence that converges in distribution.  Let $D\subset \mathbb{R}$ denote the set of continuity points of the limit distribution.  For $x\in D$, note that  $Y_N(x)$ converges in distribution (along the subsequence) by the continuous mapping theorem.  Thus, since  $Y_N(x)$ is uniformly integrable as the second moment $E Y_N^2(x) \le   E X_N^2$ is uniformly bounded by (P2), we obtain that $y_N(x)$ has a pointwise limit for all $x\in D$.   

\medskip
  Below we will show that there is a unique continuous function $y(x)$ 
 such that $y_N(x)\to y(x)$ for all $x\in D$.  Then,  using the monotonicity of $y_N(x)$ over either  $x\ge 0$ or $x\le 0$, the whole sequence $y_N(x)$ must converge pointwise to $y(x)$ for all $x\in \mathbb{R}$.  The functions $y_N(x)$ are  right-continuous, so, by the same argument that is used to prove the Glivenko--Cantelli theorem and using the continuity of $y(x)$, we will  also then have uniform  convergence $y_N(x)\to y(x)$ for $x\in \mathbb{R}$, as claimed.  
\medskip

We have shown that $y(x)=\lim_{N\to \infty} y_N(x)$ exists  for  $x\in D$ (a dense subset of $\mathbb{R}$),  when the limit is taken  over a  subsequence of $y_N$.  Now extend the definition of $y(x)$ to a general $x \in \mathbb{R}$ by taking a sequence $z_r\in D$ such that $z_r\downarrow x$ and  setting  
$y(x)\equiv\lim _{r\to \infty} y(z_r)$ . 
Since $y(x)$ shares the same monotonicity properties as the limit of $y_N(x)$ on $x\in D$, it is well-defined, i.e., not dependent on the choice of the sequence $z_r$.  Note  that $y(x)$ is  right-continuous (by construction), and $y_N(x)\to y(x)$  for all $x\in \mathbb{R} $ at which $y(x)$ is continuous.  In particular,  $y_{N}(x)\to y(x)$ a.e.\ $[dx]$, since $y(x)$ has at most countably many discontinuities.

\medskip
  From the recursion relation (\ref{rec}) we have
$${y_{N}(x_{N,n+1})-y_{N}(x_{N,n})\over x_{N,n+1} -x_{N,n}} ={x_{N,n+1}/(N-1)\over -S_{N,n}^{-1}}= {x_{N,n}-S_{N,n}^{-1}\over - S_{N,n}^{-1} (N-1)} = {1\over N-1}-x_{N,n}y_{N}(x_{N,n}).$$ 
Let $0\le u<v$ be continuity points of  $x\mapsto y(x)$, and take  $N$ to be sufficiently large that $x_{N,1}> v$, so $n(v,N)$ and $n(u,N)$ are defined.   Multiply the first and last parts of the above display by $x_{N,n+1} -x_{N,n}$ and sum
 over $n$ from $n(v,N)$ to $n(u,N)$, to obtain an equation of the form $A_N=B_N-C_N$.   Here $A_N$ and $B_N$ are  telescoping sums:  
 $$A_N= y_N(u)-y_N(v) \to y(u)-y(v), \ \  B_N =(u_N-v_N)/(N-1)\to 0,$$ where $u_N=x_{N,n(u,N)}$, $v_N=x_{N,n(v,N)}$, and 
 $$C_N = \int_{u_N}^{v_N} h_N(x) y_N(x)\, dx,$$
where $h_N(x)\equiv x_{N,n(x,N)}$.   Note that by  (\ref{mesh}) we have
$h_N(x)  \to x$ for all $x\in \mathbb{R}$.  
 This leads to the integral equation 
$$y(u)-y(v)=-\int_u^v x y(x)\, dx$$
by applying the  bounded convergence theorem, since $h_N(x)y_{N}(x)\to xy(x)$  a.e.\ $[dx]$; note that  the $y_N(x)$ are uniformly bounded, since they are nonnegative, unimodal, and converge pointwise.  Moreover, by the right-continuity of $y(x)$, the integral equation holds for all $0\le u<v$, and,    
 by the symmetry property the case  $ u<v\le 0$ is also  covered by the above argument.   Therefore $y(x)$  is differentiable and satisfies the linear first-order ODE $y'(x) +xy(x) =0$.  This ODE has general solution of the form $y(x)=c\, \varphi(x)$, where $c$ is a constant and $\varphi(x)$ is the standard normal density.

\medskip
It remains to identify $c$.  By  (\ref{den}), \begin{equation} \int_{-\infty}^\infty y_N(x)\, dx=\sum_{n=1}^{N-1} [(N-1) (x_{N,n} -x_{N,n+1})]^{-1} (x_{N,n} -x_{N,n+1} )= 1,\label{int}\end{equation}\medskip
so by Fatou's lemma (applicable since $0\le y_N(x)\to y(x)$ a.e.)
$$c=\int_{-\infty}^\infty y(x)\, dx \le \liminf_{N\to \infty} \int_{-\infty}^\infty y_N(x)\, dx =1.$$
Let $\tilde X_N$ be a random variable having density $y_N(x)$. The second moment of $\tilde X_N$ \begin{eqnarray*}
\int_{-\infty}^\infty x^2\,y_N(x)\, dx &=&\sum_{n=1}^{N-1} [(N-1) (x_{N,n} -x_{N,n+1})]^{-1} (x_{N,n}^3 -x_{N,n+1}^3)/3\\
&\le &(N-1)^{-1} \sum_{n=1}^{N-1} x_{N,n}^2 \le  1,
\end{eqnarray*}
where the last inequality follows from the variance bound (P2) in Lemma \ref{lem1}. 
This implies that the $\tilde X_N$ are tight, so for any $\epsilon>0$ there exist $u<v$ such that 
$\int_u^v y_N(x)\, dx > 1-\epsilon$ for all $N$.
By the bounded convergence theorem,  $$\int_u^v y_N(x)\, dx \to \int_u^v y(x)\, dx = c\int _u^v \varphi(x)\, dx,$$ 
so  $\int_u^v y_N(x)\, dx <c+ \epsilon$ for $N$ sufficiently large.  This shows that $c> 1-2\epsilon$, but as $\epsilon >0$ was arbitrary and $c\le 1$, we have established that $c=1$.  

\medskip
This uniquely identifies the function $y(x)$ as $\varphi(x)$, so we have shown that
$y_{N}(x)\to y(x)=\varphi(x)$ for all $x\in \mathbb{R} $.  Hence the distribution of $\tilde X_N$ (having density $\tilde p_N$) converges in total variation distance (and consequently in distribution) to ${\cal N}(0,1)$.  
The last part of the proof shows that it is possible to create a coupling  between  $X_N\sim \mathbb{P}_N$ and $\tilde X_N$ (on the same probability space) such that  \begin{equation}
\label{coupled}
|X_N-\tilde X_N| \le |x_{N,n}-x_{N,n+1}| \hbox{  when  } \tilde X_N \in [x_{N,n+1},x_{N,n}].\end{equation}   We then  have the uniform bound \begin{equation}
\label{ucoupled} |X_N-\tilde X_N|\le \max_{n=1,\ldots,N-1}|x_{N,n}-x_{N,n+1}|= \delta_N \to 0
\end{equation} using (\ref{mesh}), so by Slutsky's lemma we conclude that $\mathbb{P}_N$ converges in distribution to ${\cal N}(0,1)$.  

\medskip
To create the above coupling between $X_N$ and $\tilde X_N$, note that $\tilde X_N$ uniformly distributes mass $1/(N-1)$ on each interval between successive  $x_{N,n}$.  Let $N$ be odd (a similar argument works for $N$ even), in which case $x_{N,m}=0$ for $m=(N+1)/2$.  
We need to split each interval  in such a way that there is mass $1/N$ assigned to $x_{N,n}$ from the two adjacent parts (or the one adjacent part if $n=1$ or $N$).  
This can be done as follows.
Split the first interval to the right of zero so  there is mass $L_1 = 1/(2N)$ on the left part and $R_1=1/(N-1) - 1/(2N)$ on the right part.  
Split the $j$-th  interval to the right of zero so there is mass $L_j= (2j-1)/(2N) -(j-1)/(N-1)$ on the left part, and $R_j = 1/(N-1) - L_j = j/(N-1) -(2j-1)/(2N)$ on the right part, for $j=2,\ldots,(N-1)/2$. 
Some algebra shows that the mass assigned the right endpoint of the $j$-th interval  is $R_j + L_{j+1} = 1/N$  for $j=1,\ldots,(N-1)/2-1$, and for the last interval ($j=(N-1)/2$) it is also $1/N$.
Use symmetry to define the coupling over the negative intervals. Note that the median  $x_{N,m}=0$ gets mass $1/((2N) + 1/((2N) =1/N$  as well. 
This completes the proof of Theorem \ref{thm}. \qed

\bigskip

\section*{Application of Stein's method} Stein's method allows us to obtain bounds on the rate of convergence.  The rate will be measured using Wasserstein distance: for two probability distributions $\mu$ and $\nu$ on $\mathbb{R}$, 
$$d_W(\mu,\nu)=\sup_{h\in {\cal L}} |Eh(X)-Eh(Y)|$$
where $X\sim \mu$ and $Y\sim \nu$, and $\cal L$ is the collection of 1-Lipschitz functions $h\colon \mathbb{R} \to \mathbb{R}$ such that $|h(x)-h(y)|\le |x-y|$.  
Here we derive  bounds on $d_W(\mathbb{P}_N, {\cal N})$, where $\cal N$ is Gaussian with the same mean and variance as $\mathbb{P}_N$.

 \cite{Gold1997} introduced the notion of  zero-bias distributions, defined as follows. Given a r.v.\ $X$ with mean zero and variance $\sigma^2$, there is a r.v.\ $X^*$ such that $\sigma^2 E[f'(X^*)]=E[Xf(X)]$
for all  absolutely continuous functions $f\colon \mathbb{R}\to \mathbb{R}$ for which $E|Xf(X)|<\infty$. (This result  also appears as Proposition 2.1 of \cite{che2010}, although   a slight correction is needed: the $\sigma^2$ is misplaced in the first display).  

The distribution of $X^*$ is the $X$-{\it zero-bias distribution}.  It has  density $p^*(x)=E[XI(X>x)]/\sigma^2$.  The unique fixed point of the zero-bias transformation is ${\cal N}(0,\sigma^2)$, and the intuition behind Stein's method is that if $X$ is close to $X^*$ it should be close in distribution to ${\cal N}(0,\sigma^2)$.  Indeed, from Lemma 2.1 of \cite{Gold2004}, the Wasserstein distance between $X$ and a normal  variable  having the same mean and variance is bounded above by $2E|X-X^*|$ when $X$ and $X^*$ are defined on the same probability space.

In our setting,  $\tilde X_N$ has the $X_N$-zero-bias distribution because its density $y_N(x)$ agrees with $p^*(x)$ when $X=X_N$.  Further, we have coupled  $\tilde X_N$ and $X_N$  to satisfy (\ref{coupled}).  Thus 
\begin{equation}
\label{ratebound}
d_W(\mathbb{P}_N, {\cal N}) \le 2E|X_N-\tilde X_N|\le {2\over N-1} \sum_{n=1}^{N-1}(x_{N,n}-x_{N,n+1})=  {4x_{N,1}\over N-1}.
\end{equation}

Larry Goldstein pointed out that it is possible to obtain a lower bound on the Wasserstein distance between $\mathbb{P}_N$ and its zero-bias distribution as follows. Consider the ``sawtooth" piecewise linear function $h\colon \mathbb{R} \to \mathbb{R}$ defined to have  knots at each $x_n$ and at each midpoint $m_n =(x_n+x_{n+1})/2 $ between successive  $x_n$, such that $h(x_n)=0$ and $h(m_n)= (x_n-x_{n+1})/2$, with $h$ vanishing outside $[x_N,x_1]$.  Clearly $h$ is 1-Lipschitz and $Eh(X_N)=0$, so from its definition the Wasserstein distance between $\mathbb{P}_N$ and its zero-bias distribution is bounded below by
$Eh(\tilde X_N) =x_{N,1}/[2(N-1)]$, which is of the same order as the upper bound on $d_W(\mathbb{P}_N, {\cal N})$.  

This strongly suggests $d_W(\mathbb{P}_N, {\cal N}) \asymp x_{N,1}/N$.
Moreover,  as mentioned earlier, there is convincing numerical evidence that $x_{N,1} \asymp q_N$, where $q_N$ is the upper $1/(2N)$-quantile of ${\cal N}(0,1)$.   Indeed, we  have already shown 
$x_{N,1} \ge \sqrt{\log N}/2$, and using Mills ratio
 it can be shown that $q_N\asymp \sqrt{\log N}$, so combining with  ({\ref{ratebound}) we expect $d_W(\mathbb{P}_N, {\cal N})\asymp \sqrt{\log N}/ N$. 
In terms of what we have actually proved, however,  the  uniform coupling property (\ref{ucoupled}) and (\ref{mesh}) only give the   weaker upper bound $d_W(\mathbb{P}_N, {\cal N}) \le 2 \delta_N\le  4 / \sqrt{\log N}.$


\section*{Proof of Theorem \ref{thm2}}

 We have already shown that $X_{t}^{(m)}$ converges in distribution to the OU process $X_t$. By  appealing to Slutsky's lemma for random elements of metric spaces \cite[Theorem 18.10]{van2000}, it thus suffices to show that  for each $T>0$ the processes $\{\xbar X_{t}, t\in [0,T]\}$ and $\{X_{t}^{(m)}, t\in [0,T]\}$ can be coupled  as random elements of $D[0,T]$ on a joint probability space, with their difference tending uniformly to zero in probability.  The  upper bound on the Wasserstein distance just derived implies that if $X_N\sim \mathbb{P}_N$, there exists $Z\sim {\cal N}(0,1)$ on a joint probability space with
$E|X_N-Z|=O(1 / \sqrt{\log N})$. 

Further, any sequence of iid-$\mathbb{P}_N$ r.v.s can be coupled in this way using independent coupled pairs on a joint probability space.   The  single replacement mechanism that generates samples of size  $m$  from $\mathbb{P}_N$ can  be coupled with a chain of samples from ${\cal N}(0,1)$ by using the same randomly selected index in each transition. Each transition involves selecting the update from an independent sample of size $m$, so $m([mT]+1)$  coupled pairs are involved over the interval $[0,T]$.
Thus we have constructed a coupling of the processes $\xbar X_{t}$ and $X_{t}^{(m)}$ over $t\in [0,T]$ with 
$$E\bigg\{\sup_{t\in [0,T]}|\xbar X_{t}-X_{t}^{(m)}|\bigg\} \le m([mT]+1) E|X_N-Z|/\sqrt m=m^{3/2}O(1 / \sqrt{\log N})\to 0$$
since we have assumed $m=o({\log N})^{1/3}$, so the result follows by Chebyshev's inequality. \qed

\section*{Proof of Lemma \ref{lem1}} 

The following result is needed to prove Lemma \ref{lem1}. Denote $m=m_N=(N+1)/2$ if $N$ is odd and $m=m_N=N/2$ if $N$ is even.  For any given  $x_1>0$, let $x_2,\ldots, x_N$ be generated by the recursion (\ref{rec}).  We consider each term $x_n=x_n(x_1)$ as a function of $x_1$, and similarly consider each cumulative sum  $S_n=x_1+\ldots +x_n$ as a function of $x_1$,  $S_n = S_n(x_1)$, for $n=2,\ldots,N$.  

\medskip

\begin{lemma} For  all  $n = 2,\ldots,m $,
\label{lem2}
\begin{itemize}
\item[\rm{(a)}] There exists a unique positive real number $a_n$ such that $x_n(a_n)=0$ and for which if $x_n(z)=0$ then $z \le a_n$.  The $a_n$ are strictly increasing: $0<a_2<\ldots <a_m$.  
\item[\rm{(b)}]  For $x_1>a_n$,  $x_n$ is a positive, increasing, and continuous function of $x_1$.
\item[\rm{(c)}]  There exists a unique positive real number $b_n$ such that $S_n(b_n)=0$ and for which if $S_n(z)=0$ then $z \le b_n$.  The $b_n$ are strictly increasing: $0<b_2<\ldots <b_m$.
\item[\rm{(d)}]  For $x_1>b_n$,  $S_n$ is a positive, increasing, and continuous function of $x_1$.
\item[\rm{(e)}]  $a_n>b_n$.
\end{itemize}

\end{lemma}
\medskip

\noindent {\bf Proof}.  We use  induction on $n$.  Clearly $a_2=1$ is the unique positive solution of $x_2=x_1-x_1^{-1}=0$, and  $x_2$ is positive, increasing, and continuous in $x_1$ for $x_1>1$.  The equation $S_2 = x_1 + x_2 = 2x_1 - x_1^{-1} = 0$ has the unique positive solution $b_2 =\sqrt{2}/2 < a_2$, and $S_2$ is positive, increasing, and continuous in $x_1$ for $x_1>b_2$ since both $2x_1$ and $-x_1^{-1}$ are increasing, continuous functions of $x_1$.  Note that this holds true even though $x_2<0$ for $x_1<a_2$.  This completes the initial induction step $n=2$.

\medskip
Suppose we have determined constants $a_i$ and $b_i$ satisfying properties (a)--(e) for $i=1,\ldots,n < m$.  We show these properties hold for $i = n+1$.  First, we assert that there are values of $x_1>a_n$ such that $x_{n+1}<0$.  To see this, note that for any $x_1 > a_n$, we have $x_1>a_n>b_n\ge b_i$ for $i=1,\dots ,n$.  Thus $S_i>0$, which implies $x_{i+1}=x_i-S_i^{-1}<x_i$, hence $S_n<nx_1$.  Then $x_{n+1}=x_n-S_n^{-1}<x_n-(nx_1)^{-1}<x_n-(2na_n)^{-1}$ for $x_1$ sufficiently close to $a_n$, e.g., $a_n<x_1<2a_n$.  But $x_n$ can be made arbitrarily close to zero for $x_1$ sufficiently close to $a_n$ by continuity, in particular  $x_n<(2na_n)^{-1}$, from which it follows $x_{n+1}<0$.

\medskip
Next, we assert there are values of $x_1>a_n$ such that $x_{n+1}>0$.  Note that for such $x_1$, each $x_i>0$ by property (a) and (b), so $S_n>x_1$.  Thus $x_{n+1}=x_n-S_n^{-1}>x_n-x_1^{-1}$.  As $x_1$ becomes sufficiently large, $x_n$ remains bounded away from zero while $-x_1^{-1}$ can be made arbitrarily close to zero.  It follows that for sufficiently large $x_1>a_n$ we have  $x_{n+1}>0$.

\medskip
Thus for $x_1>a_n>b_n$, by continuity of $x_n$ and $S_n>0$ as functions of $x_1$ under the inductive hypothesis, $x_{n+1} =x_n-S_n^{-1}$  is continuous, so the intermediate value theorem implies the existence of at least one root of the equation $x_{n+1}(x_1)=0$, and that root is strictly greater than $a_n$.  The argument of the preceding paragraph showed that the set of roots of $x_{n+1}=0$ is bounded from above, so we determine $a_{n+1}$ uniquely as the supremum of the non-empty, bounded set $\{x_1>a_n: x_{n+1}(x_1)=0\}$, and that supremum satisfies $a_{n+1}> a_n> b_n$.  In fact, the set is finite because the equation $x_{n+1}(x_1)=0$ is equivalent to a polynomial equation with finitely many real roots, so  we can say ``maximum" rather than ``supremum".  It is then clear that $x_{n+1}(a_{n+1})=0$.  Then $x_{n+1}=x_n-S_n^{-1}$ is an increasing, continuous function for $x_1>a_{n+1}$ because both $x_n$ and $-S_n^{-1}$ are increasing and continuous, and so $x_{n+1}$ is a positive, increasing, and continuous function of $x_1$ for $x_1>a_{n+1}$.  This establishes parts (a) and (b) of the inductive step.

\medskip
Next, we establish parts (c) and (d) of the inductive step.  For $x_1$ greater than but sufficiently close to $b_n$, $x_{n+1}=x_n-S_n^{-1}$ can be made arbitrarily large negative, because $x_n$ approaches the constant $x_n(b_n)$ while $S_n$ goes to zero from above.  Therefore $S_{n+1}=x_{n+1}+S_n$ also becomes arbitrarily large negative as $x_n$ approaches $b_n$ from above.  Writing
 $$S_{n+1}=S_n+x_{n+1}=S_n+x_n-S_n^{-1}=S_n+x_{n-1}-S_{n-1}^{-1}-S_n^{-1}=\ldots =S_n+x_1-x_1^{-1}-S_2^{-1}-\ldots
-S_n^{-1}$$
we find that $S_{n+1}$ is continuous and increasing for $x_1>b_n>\ldots >b_2$, because each term in the sum on the right-hand side is continuous and increasing for such $x_1$ by the inductive hypothesis.  Furthermore, we have established that both $x_{n+1}$ and $S_n$ are positive for $x_1$ sufficiently large, so for such $x_1$, $S_{n+1}=x_{n+1}+S_n$ is also positive.  Thus by the intermediate value theorem, there exists a root of the equation $S_{n+1}(x_1)=0$ strictly greater than $b_n$.  Since the set of such roots is bounded from above, we define $b_{n+1}$ uniquely as the maximum of the non-empty, bounded, finite set $\{x_1>b_n\colon  S_{n+1}(x_1)=0\}$, the maximum of which satisfies $b_{n+1}>b_n$ and $S_{n+1}(b_{n+1})=0$.  We conclude that $S_{n+1}$ is a positive, increasing, continuous function of $x_1$ for $x_1>b_{n+1}>b_n$.  This establishes properties (c) and (d) of the inductive step.

\medskip
To establish property (e), argue by contradiction.  We have already shown that  $a_{n+1}>a_n$ and $a_n>b_n$ by the inductive hypothesis.  At $x_1=b_{n+1}$, we have $0=S_{n+1}= x_{n+1}+S_n$, i.e., $x_{n+1}=-S_n$.  So suppose it were the case that  $b_{n+1} \ge a_{n+1}$.  Then $x_{n+1}$ would be strictly negative, because we would have $x_1=b_{n+1} \ge a_{n+1}> a_n > b_n$, so that $S_n>0$ by the inductive hypothesis.  But if $x_1>a_{n+1}$, then $x_{n+1}<0$ contradicts property (a), which states that $x_{n+1}$ is positive for such $x_1$, or if $x_1=a_{n+1}$, then $x_{n+1}<0$ contradicts the defining property of $a_{n+1}$, namely, $x_{n+1}=0$.  Thus $a_{n+1}>b_{n+1}$.  This establishes part (e) of the inductive step, and the proof of Lemma \ref{lem2} is complete. \qed

\bigskip
\noindent{\bf Proof of Lemma \ref{lem1} (continued)}.  First consider the case that $N$ is odd, and set $m=(N+1)/2$.  
To prove the symmetry property (P3), we need to show that if $x_1$ is any root of the equation $x_m(x_1)=0$, then  the identity $x_{m+i}=-x_{m-i}$ holds for $i=0,1,\ldots, m-1$.  The proof is by induction on $i$.  The case $i=0$ is immediate  as it is simply $x_m=0$.  Suppose the identity  holds up to a given index $i<m$.  Then
\begin{equation*}
x_{m+i+1}=x_{m+i}-S_{m+i}^{-1}=-x_{m-i}-S_{m-i-1}^{-1}=-(x_{m-i-1}-S_{m-i-1}^{-1})-S_{m-i-1}^{-1}=-x_{m-i-1},
\end{equation*}
where the second equality is by the inductive hypothesis, since the symmetry $x_{m+i} = -x_{m-i}$ for values of the subscript $m,\ldots,m+i$ on the left and $m-i$ on the right also implies that $S_{m+1} = S_{m-i-1}$. The first and third equalities are by the recursion, so the identity holds for $i+1$, and we have shown (P3).  The zero-mean property (P1) follows from the symmetry (P3).  For the variance property (P2), denoting $S_0=0$ we have
\begin{eqnarray*}
N-1&=&\overset{N-1}{\underset{n=1}{\sum }}S_nS_n^{-1}=\overset{N-1}{\underset{n=1}{\sum
}}S_n(x_n-x_{n+1})=\overset{N-1}{\underset{n=1}{\sum }}[(S_{n-1}+x_n)x_n-S_nx_{n+1}]\\
&=&\overset{N-1}{\underset{n=1}{\sum
}}[S_{n-1}x_n-S_nx_{n+1}+x_n^2]\\
&=& x_1^2+\ldots
+x_{N-1}^2-S_{N-1}x_N,
\end{eqnarray*}
where we used the recursion in the second equality, and the last equality is from a telescoping sum.   (P1) implies $-S_{N-1}=x_N$, and (P2) follows.  

\medskip

The proof of  the second part of the lemma relies on the  

\medskip  {\it Claim}: There exists a unique zero-median solution $x_1,\ldots, x_N$
that maximizes $x_1$ in the sense that if  $\tilde x_1,\ldots, \tilde x_N$ is any other zero-median solution then $\tilde x_1 < x_1$, and this solution satisfies (P4).

\medskip
To prove this claim when $N$ is odd, let $x_1=a_m>0$ provided by Lemma \ref{lem2} (a) in the special case $n=m$, so that $x_m(x_1)=0$ (i.e., the zero-median property holds) and $x_1$ is the largest possible root of $x_m=0$, establishing the existence and uniqueness parts of the claim.   For property (P4), by Lemma \ref{lem2} we have that  $x_1=a_m>b_m>\ldots >b_2$, and $S_n>0$ for  $n=1,\ldots,m$, so $x_n-x_{n+1}=S_n^{-1}>0$ for those $n$.  The zero-median property and the symmetry then imply $x_n>x_{n+1}$ for the remaining  $n=m+1,\ldots,N$, so (P4) holds. 

\medskip
Next consider the case that $N$ is even, and set $m=N/2$.  The symmetry property (P3) follows by a similar inductive argument on $i$ to what we used earlier,  so the zero-mean property (P1) also  holds.   (P2) was proved earlier without using any restriction on $N$. 
To prove the  claim, we
need to show that there is a largest root, call it $a_{m+\frac12}$, of the equation  $x_m(x_1)+x_{m+1}(x_1) = 0$. 
The sum on the left is $2x_m-S_m^{-1}$, which by Lemma \ref{lem2} (b) and (d) is continuous for $x_1>b_m$, negative at $x_1= a_m$, and positive for sufficiently large $x_1>a_m$.  Thus there is a root of the above equation greater than $a_m$. Define $a_{m+\frac12}$ as the unique maximum of the non-empty, bounded, finite set of roots. 
Taking $x_1= a_{m+\frac12}$ establishes the existence of a solution to the recursion having the zero-median property, as well as its uniqueness in maximizing $x_1$. 
For the property (P4) that the resulting sequence is strictly decreasing, note that by Lemma \ref{lem2} we have $x_1=a_{m+\frac12}>a_m>b_m>\cdots >b_2$, and $S_n>0$ for  $n=1,\ldots,m$, so $x_n-x_{n+1}=S_n^{-1}>0$ for those $n$.  The zero-median property and the symmetry then imply $x_n>x_{n+1}$ for the remaining $n=m+1,\ldots,N$, completing the proof of the claim.

\medskip  To complete the proof of the lemma, it remains to show uniqueness of zero-mean monotonic sequences $x_1 > x_2 >\ldots > x_N$  generated by the recursion. The zero-mean condition is the same as $S_N(x_1)=0$. 
 The key point is that with $x_1>0$, assuming we have a monotonic solution implies that  the cumulative sums $S_n$ are positive for $n=1,\ldots,N-1$, by the recursion equation $x_{n+1} = x_n - S_n^{-1}$.   Argue by induction as follows.  Clearly $x_2 = x_1 -1/x_1$  is an increasing function of $x_1>0$, even if $x_2$ happens to be negative.  Therefore $S_2 = x_2 + S_1$ is an increasing function of $x_1$ and thus a positive increasing function for $x_1$ greater than the given $x_1$ by monotonicity.  Therefore $x_3 = x_2-S_2^{-1}$ is an increasing function of $x_1$ (even if it is negative).  Therefore $S_3 = x_3 + S_2$ is an increasing function of $x_1$ and thus a positive increasing function for $x_1$ greater than the given $x_1$ by monotonicity.  And so on, so by induction we have $x_{N}$ is an increasing function of $x_1$ and $S_{N-1}$ is a positive increasing function of $x_1$ for $x_1$ greater than the given $x_1$ by monotonicity.  Therefore $S_N = x_{N}+S_{N-1}$ is increasing in $x_1$ for $x_1$ greater than the given $x_1$ by monotonicity.
   If we assume the given $x_1$ is smaller than the $a_m$  constructed in proving the  claim (when $N$ is odd, similarly when $N$ is even), then we must have $S_N(a_m)>0$, which contradicts the zero-mean property.   The same argument shows there is no  $x_1>a_m$ that generates a monotonic solution with the zero-mean property, and we conclude that $a_m$ generates the  unique zero-mean monotonic solution.
 \qed

\section*{Derivation of the recursion}

In the ground state, the Hamiltonian  depends only on the locations of the particles ${\bf x} =(x_1,\ldots , x_N)$,  $x_1>x_2>\ldots > x_N$:  $H({\bf x}) = V({\bf x}) + U({\bf x})$,
where 
$V({\bf x}) = \sum_{n=1}^N x_n^2$
is the classical potential for $N$ (non-interacting) particles of 
equal mass in a parabolic trap, and
$$U({\bf x}) = \sum_{n=1}^N\left({1\over x_{n+1}-x_n} -{1\over x_{n}-x_{n-1}}\right)^2$$
is the hypothesized ``interworld" potential, where
$x_0=\infty$ and $x_{N+1}=-\infty$.  Write

\begin{eqnarray*}
(N-1)^2 
&=& \left[ \sum_{n=1}^{N-1} {  x_{n+1}-x_n \over x_{n+1}-x_n}\right]^2 \\
&=& \left[  \sum_{n=1}^{N} \left(  {1\over  x_{n+1}-x_n}- {1\over  x_{n}-x_{n-1}} \right) (x_n-\bar x_N)  \right]^2\\ 
&\le&  \sum_{n=1}^N\left({1\over x_{n+1}-x_n} -{1\over x_{n}-x_{n-1}}\right)^2 \sum_{n=1}^{N} (x_n-\bar x_N)^2\\
& \le& U({\bf x})\, V({\bf x}), \end{eqnarray*}
where the first inequality is Cauchy--Schwarz.  
So $U\ge (N-1)^2/V$, leading to  
$$H=U+V \ge (N-1)^2/V +V \ge 2(N-1)$$
with the last inequality being equality for $V= N-1$.
We conclude that $\bf x$ is a ground state solution if and only if (P1) and (P2) hold,
and
$$x_n= {\alpha \over x_{n+1}-x_n} -{\alpha\over x_{n}-x_{n-1}}$$
for some constant $\alpha$.
The sum of the right of the above display telescopes, leading to the recursion (\ref{rec})
by rearranging and noting that $\alpha =-V/(N-1) =-1$ by a similar argument to the proof of (P2) in Lemma 1.

\section*{Acknowledgements}
The authors (IM and BL) are grateful to Larry Goldstein and Adrian R\"ollin for many stimulating discussions and helpful comments.  
The research of IM was partially supported by NSF Grant DMS-1307838 and NIH Grant 2R01GM095722-05, and that of BL by NIH Grant P30-MH43520.  IM  thanks 
the Institute for Mathematical Sciences at National University of Singapore for support during the {\it Workshop on New Directions in Stein's Method} (May 18--29, 2015) where the paper was presented.

\bibliographystyle{elsarticle-harv} 
\bibliography{clt-relativity}

\end{document}